\documentclass[10pt]{dis03}
\usepackage[american]{babel}
\usepackage{epsf,amsmath}
\newcommand{\by}[2]{\leavevmode\kern.1em
\raise.5ex\hbox{\ensuremath{\scriptstyle #1}}\kern-.1em
/\kern-.15em\lower.25ex\hbox{\ensuremath{\scriptstyle #2}}}
\newcommand{\eVdist}{\kern-0.06667em}
\newcommand{\gev}{{\,\text{Ge}\eVdist\text{V\/}}}
\begin{document}
\title{Inelastic photo- and electroproduction of charmonium}
\author{{\scshape Igor Katkov}\\[3mm]
on behalf of the ZEUS Collaboration\\[3mm]
Skobeltsyn Institute of Nuclear Physics\\
Moscow State University\\
E-mail: katkov@mail.desy.de\\[5mm]}
\maketitle
\begin{abstract}
\noindent%
Measurements of inelastic production of charmonium with the 
ZEUS detector at HERA are presented. $J/\psi$ and $\psi'$ mesons 
have been identified using the decay channel 
$\psi\to\mu^{+}\mu^{-}$.
The data, corresponding to an integrated luminosity of $38\,\text{pb}^{-1}$ in
photoproduction and $73.3\,\text{pb}^{-1}$ in electroproduction,
are confronted to theoretical predictions.
\end{abstract}
\section{Introduction}
Inelastic production of charmonium involves
two stages: the creation of a heavy quark pair at short distances
and the subsequent formation of the $\psi$ bound state
which occurs at long-distance scales. 
In the inelastic process $e\,p\to e\,\psi\,X$, 
the $c\bar{c}$ pair production is 
dominated by photon-gluon fusion,
$\gamma^{*}g\to c\bar{c}$, and can be calculated in perturbation theory. 
The $\psi$ bound state can be considered to be formed
by a $c\bar{c}$ pair in either a colour singlet (CS) or colour
octet (CO) state.
In the colour singlet model (CSM) only CS contribution is assumed.
In the framework of non-relativistic QCD (NRQCD) both CS and CO
contributions exist and the latter contribution is parametrised
using a set of long distance matrix elements tuned to describe
hadroproduction data.

\begin{figure}[!b]
\vspace*{5.7cm}
\begin{center}
\includegraphics{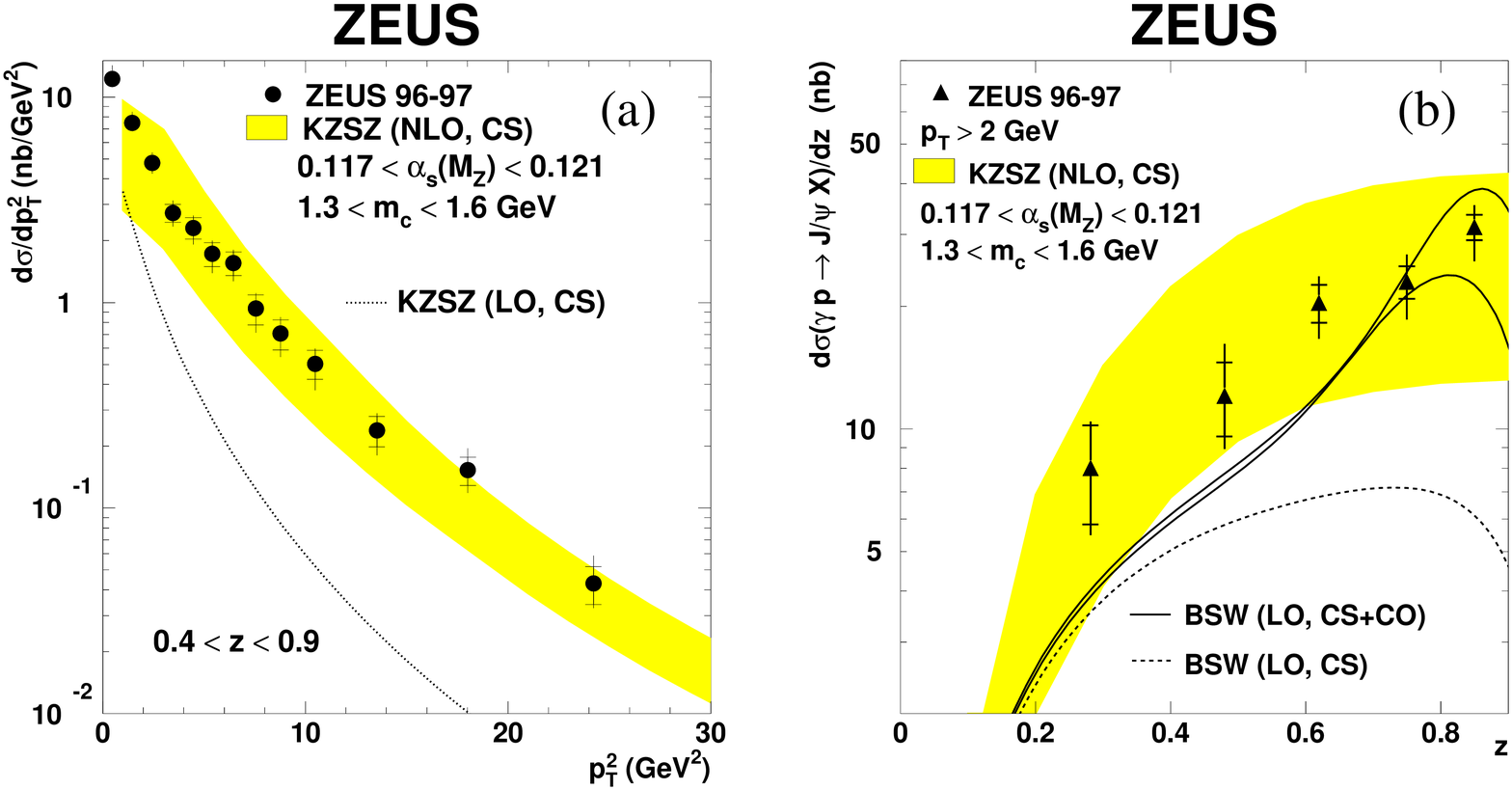}
\end{center}
\vspace*{-0.5cm}
\caption{Photoproduction of $J/\psi$ mesons.
Differential cross sections as a function of (a) transverse 
momentum squared and (b) inelasticity. The data are compared to
an NLO CSM calculation (shaded band), where the spread is due to 
uncertainties on $m_c$ and $\Lambda_\text{QCD}$, (a) LO CSM 
calculation (dotted line) and (b) LO NRQCD calculation with 
soft gluon emission resummation,
including CS terms (dashed line) and sum of CS and CO terms 
(solid lines), where the spread is due to the uncertainty on 
the shape-function parameter.}\label{fig:1}
\end{figure} 

In the semi-hard or $k_T$-factorisation 
approach, based on non-collinear
parton dynamics governed by the BFKL
evolution equations, effects of non-zero initial 
gluon virtuality (transverse momentum) are taken into account.

The inelasticity, $z$, which is the fraction
of the virtual photon energy transferred to the $\psi$ in the 
proton rest frame, is sensitive to the various production 
mechanisms. CS processes are expected to contribute to the region of 
medium $z$ values,  whereas CO (and diffractive) processes
populate the high-$z$ region. Resolved-photon processes, in which the photon 
acts as a source of incoming partons, populate low values of $z$.
\section{Charmonium photoproduction}
In photoproduction, charmonium production was measured by ZEUS~\cite{ZEUS:02} in 
the kinematic range
$50<W<180\gev$, where $W$ is the photon-proton centre-of-mass energy,
and $0.2<z<0.9$. 

The $\psi^\prime$ to $J/\psi$ cross-section ratio was
also measured and used to estimate the fraction of $J/\psi$ mesons coming
from $\psi^\prime$ cascade decays. This value is about $15\%$, consistent
with expectations. 
Hence all presented theoretical predictions of the $J/\psi$ differential 
cross sections were scaled by $1.15$.

In Fig.~\ref{fig:1}(a), comparison of the data with the 
CSM calculation~\cite{Kraemer:CSM} 
in LO and NLO (available only for the direct photon process) 
is shown. The $p_T^2$ 
spectrum in LO (KZSZ(LO, CS)) is considerably softer
than one observed in the data. The NLO calculation (KZSZ(NLO, CS)) 
describes well both the shape and normalisation
of the data, although theoretical uncertainties are large. 
        
The NLO calculation also describes well the inelasticty distribution shown in 
Fig.~\ref{fig:1}(b). 
Kinematic limitations important at phase space boundaries (high $z$ values)
were considered~\cite{BSW:00} in the framework of NRQCD
(BSW(LO, CS) and BSW(LO, $\text{CS}+\text{CO}$))
and a resummation procedure
was introduced. The resummation leads to a good agreement of the sum of 
CS and CO contributions with the data at high $z$. 
At lower $z$ values the discrepancy is likely to be due to resolved 
contributions, which were not taken into account in the
calculation. The CS contribution alone is below the data.
\section{Electroproduction of $J/\psi$}
\begin{figure}[p]
\vspace*{12.0cm}
\begin{center}
\includegraphics{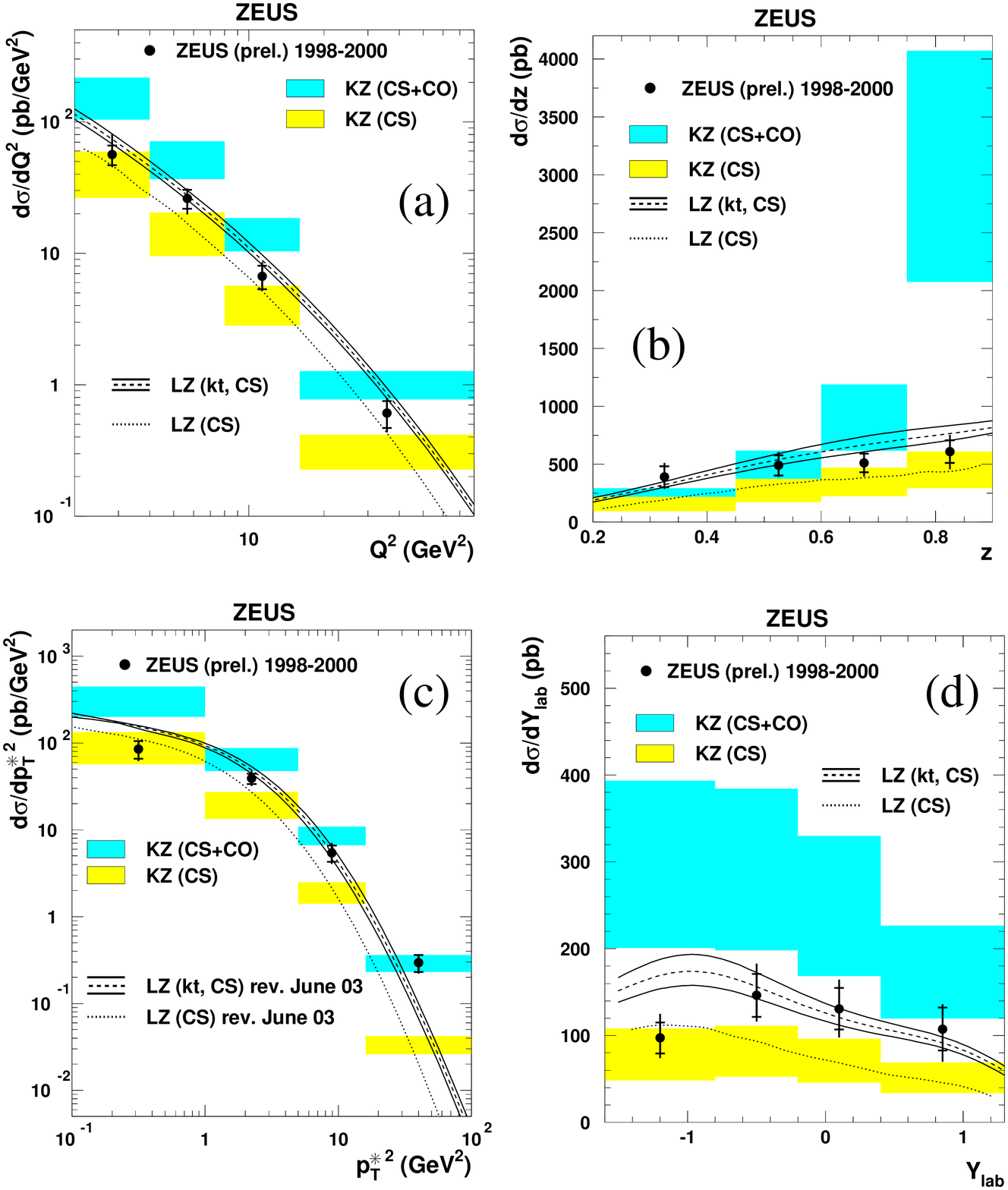}
\end{center}
\caption{Electroproduction of $J/\psi$ mesons. Differential cross sections
as a function of (a) virtuality of the exchanged photon,
(b) inelasticity, (c) transverse momentum squared 
in the photon-proton centre-of-mass system
and (d) rapidity in the laboratory system. 
The data are compared to LO NRQCD predictions~\cite{Kniehl:DIS}
(the upper band is the sum $\text{CS}+\text{CO}$ and the lower band is CS only)
and calculations in the $k_T$-factorisation approach within the CSM~\cite{Zotov:03}
(solid lines delimit the uncertainty, dashed line shows the central value).
The CSM prediction in the framework of collinear factorisation~\cite{Zotov:03}
is also shown (dotted line).}\label{fig:2}
\end{figure}

The measurement of inelastic electroproduction of $J/\psi$ was performed 
in the kinematic range $2<Q^2<80\gev^2$, $50<W<250\gev$,
$0.2<z<0.9$ and $-1.6 < Y_\text{lab} < 1.3$, where $Y_\text{lab}$ is the
rapidity of $J/\psi$ in the laboratory frame. 
In Fig.~\ref{fig:2}, the data are compared to 
predictions in the framework of NRQCD~\cite{Kniehl:DIS} 
and in the CSM with 
$k_T$-\hskip0pt factorisation~\cite{Zotov:03}.

The CS contributions (KZ(CS)) and the sum of CS and CO 
contributions (KZ($\text{CS}+\text{CO}$)) 
of the NRQCD predictions are shown separately. The uncertainty 
shown for the theoretical calculation corresponds to variations of 
the charm quark mass
($m_c=1.5\pm0.1\gev$) and the renormalisation and factorisation scales 
(from $\by12\sqrt{\smash[b]{Q^2+M^2_\psi}}$ to 
$2\sqrt{\smash[b]{Q^2+M^2_\psi}}$).
The uncertainty on the long distance matrix elements and the effect 
of different choices of parton density functions are also taken into account.
The band shows all the uncertainties added in quadrature.

In general, the CS predictions are below the data 
but consistent both in shape and 
normalisation within the uncertainties shown.
However, the prediction for the differential cross section as
a function of transverse momentum
squared in the laboratory system (not shown) and in the photon-proton 
centre-of-mass system
($p_T^{*2}$) are too soft compared to the data.
The results in the photoproduction regime,
presented in the previous section, indicate
that NLO corrections are needed to describe the $J/\psi$ transverse momentum 
spectrum in the framework of the CSM.

Inclusion of CO contributions leads to an excess of the NRQCD predictions over
the data, especially at high $z$. At high values 
of $p_T$ and $p^*_T$ agreement with the data is reasonable,
however at low values of transverse momenta the predictions overshoot
the data.

For the prediction within the $k_T$-factorisation approach 
(LZ(kt, CS)) only 
one source of uncertainty was considered, namely a variation 
of the pomeron intercept $\Delta$ which controls the normalisation 
of the unintegrated gluon density 
used in the calculation in the form suggested
by Bl\"umlein~\cite{JB}. 
Central values were calculated with $\Delta=0.35$
and the uncertainty corresponds to the variation of $\Delta$ between
$0.20$ and $0.53$.
The charm quark mass used is $m_c=1.55\gev$. The 
renormalisation and factorisation scales were set to the absolute
value of the initial gluon transverse momentum. 
A calculation in the framework of collinear factorisation
within the CSM (LZ(CS)) was also provided~\cite{Zotov:03} 
which is generally consistent with CS 
contributions of the NRQCD predictions (KZ(CS)).

Calculations based on the $k_T$-factorisation approach give a reasonable 
description of the data both in normalisation and shape. 
However, the predicted $p_T^{*2}$ spectrum 
is softer than in 
the data. 
\section{Summary}
Results on inelastic charmonium production in both photo- and electroproduction 
have been obtained with the 
ZEUS detector and
compared to theoretical predictions in the CSM,
calculations in the framework of NRQCD and the $k_T$-factorisation approach.

In photoproduction, NLO corrections to the direct photon process
calculated in the colour singlet
model describe well
the $J/\psi$ transverse momentum distribution. Cross sections 
as a function of $W$ and $z$ are also well described. 
However the theoretical 
uncertainties are large and hence CO contributions cannot be excluded.

In electroproduction, LO CS predictions, based on
collinear perturbative QCD, are below but
consistent with the data within the uncertainties, except at high
$p_T$. The LO NRQCD predictions,
including both CS and CO contributions, are generally above the data,
especially at large $z$ and small $p_T^*$ values. 
At high transverse momenta, agreement
with the data is reasonable,
but NLO corrections are needed to draw stronger conclusions.
The calculation in the $k_T$-factorisation approach within the colour
singlet model gives an overall better description of the data.

\end{document}